\documentclass{elsart}
\usepackage{graphicx}
\usepackage{epsfig,psfig}
\def\gsim{\;\raise0.3ex\hbox{$>$\kern-0.75em\raise-1.1ex\hbox{$\sim$}}\;}
\def\lsim{\;\raise0.3ex\hbox{$<$\kern-0.75em\raise-1.1ex\hbox{$\sim$}}\;}
\newcommand{\be}{\begin{equation}}
\newcommand{\ee}{\end{equation}}
\newcommand{\bea}{\begin{eqnarray}}
\newcommand{\eea}{\end{eqnarray}}
\newcommand{\bt}{\begin{tabular}}
\newcommand{\et}{\end{tabular}}
\newcommand{\ba}{\begin{array}}
\newcommand{\ea}{\end{array}}
\newcommand{\ov}{\overline}

\begin{document}

\begin{frontmatter}
\title{{\hfill \small DSF-38-2001, astro-ph/0112384}\\ $~$\\
Radiative corrections to neutrino energy loss rate in stellar interiors}
\author{S.~Esposito},
\author{G.~Mangano},
\author{G.~Miele},
\author{I.~Picardi}, and
\author{O.~Pisanti}
\address{Dipartimento di Scienze Fisiche, Universit\`{a} di Napoli
{Federico II} and INFN, Sezione di Napoli, Complesso Universitario di
Monte S. Angelo, Via Cintia, I-80126 Naples, Italy}

\begin{abstract}
We consider radiative electromagnetic corrections, at order $\alpha$, to
the process $e^+ e^- \rightarrow \nu \overline{\nu}$ at finite density and
temperature. This process represents one of the main contributions to the
cooling of stellar environments in the late stages of star evolution. We
find that these corrections affect the energy loss rate by a factor $(-4
\div 1) \%$ with respect to the tree level estimate, in the temperature and density
ranges where the neutrino pair production via $e^+ e^-$ annihilation is the
most efficient cooling mechanism.
\end{abstract}
\end{frontmatter}

\section{Introduction}
The late stages of star evolution are strongly influenced by neutrino
emission processes. After hydrogen burning, stars with small mass, of the
order of the solar mass, evolves towards a white dwarf configuration. On
the other hand more massive stars undergo several burning phases till the
formation of a central iron core and the eventual triggering of the
gravitational instability leading to the Supernova phenomenon. In both
cases the cooling rate is largely dominated by neutrino production. For
example, from the $^{12}C - ^{24}Mg$ burning phase, almost $100 \%$ of the
energy flux is emitted via neutrino production, which leave the stellar
system without any interaction because of their extremely large mean free
path. A precise determination of neutrino emission rates is therefore a
crucial issue in any careful study of the final branches of star
evolutionary tracks. In particular, changing the cooling rates at the very
last stages of massive star evolution may sensibly affect the evolutionary
time scale and the iron core configuration at the onset of the Supernova
explosion, whose triggering mechanism is still lacking a full theoretical
understanding \cite{janka}.

The energy loss rate due to neutrino emission, hereafter denoted by $Q$,
receives contribution from both weak nuclear reactions and purely leptonic
processes. However for the rather large values of density and temperature
which characterize the final stages of stellar evolution, the latter are
largely dominant. They are mainly due to four possible interaction
mechanisms:
\begin{itemize}
\item[i)] pair annihilation $~~~~~~~~~~~~~~e^+ \, + \, e^- \; \rightarrow
\; \nu \, + \, \ov{\nu}$
\item[ii)] $\nu$-photoproduction $~~~~~~~~~~~~~ \gamma \, + \, e^{\pm} \;
\rightarrow \; e^{\pm} \, + \, \nu \, + \, \ov{\nu}$
\item[iii)] plasmon decay $~~~~~~~~~~~~~~~~~~~~~~~~~\gamma^\ast \;
\rightarrow \; \nu \, + \, \ov{\nu}$
\item[iv)] bremsstrahlung on nuclei $~~~~e^{\pm} \, + \, Z \; \rightarrow
\; e^{\pm} \, + \, Z \, + \, \nu \, + \, \ov{\nu}$
\end{itemize}
Actually each of these processes results to be the dominant contribution to
$Q$ in different regions in a density-temperature plane. For very large
core temperatures, $T \gsim 10^9 \,^\circ K$, and not too high values of
density, pair annihilations are most efficient, while $\nu$ photoproduction
gives the leading contribution for $10^8 \,^\circ K \lsim T \lsim$
$10^9\,^\circ K$ and relatively low density, $\rho \lsim 10^5$ g cm$^{-3}$.
These are the typical ranges for very massive stars in their late
evolution. Finally, plasmon decay and bremsstrahlung on nuclei are mostly
important for large ($\rho \gsim 10^{6}$ g cm$^{-3}$) and extremely large
($\rho \gsim 10^{9}$ g cm$^{-3}$) core densities, respectively, and
temperatures of the order of $10^8 \,^\circ K \lsim T \lsim 10^{10}
\,^\circ K$. Such conditions are typically realized in white dwarfs.

Starting from the first calculation of neutrino energy loss rates, based on
the V-A theory of weak interactions \cite{Beaudet67a,Beaudet67b}, a
systematic study of processes i)-iv) has been carried out in a long series
of papers \cite{Beaudet67a}-\cite{Itoh96}. In all these analyses the pair
production rate i) has been evaluated at order $G_F^2$, i.e. at the zero
order in the electromagnetic coupling constant $\alpha$ expansion. Despite
of the different topology and phase space volume which characterize the
remaining processes ii)-iv), it is nevertheless worth noticing that they
are instead at least of order $\alpha G_F^2$. In view of this, it is
therefore meaningful to investigate whether including QED radiative
corrections to pair annihilation rate i) may lead to a sensible change in
the cooling rate $Q$. This is the aim of this letter. In particular we
report here the results of a calculation at order $\alpha$ of two classes
of contributions, due to vacuum radiative corrections as well as those
arising from the interaction of the $e^+$-$e^-$ pairs with the surrounding
electromagnetic plasma. This has been performed in the real time formalism
for finite temperature field theory. We actually find that the total
corrections to $Q$ range in the interval $(-4 \div 1) \%$ of the tree level
estimate, in the temperature and density ranges where the pair annihilation
process represents the main contribution to the total energy loss rate. We
have also re-evaluated both photoproduction and plasmon decay rates, and we
do find a good agreement with the most recent estimates reported in the
literature \cite{Raffelt95,Itoh96}.

The paper is organized as follows. Section 2 is a short review of the
lowest order calculation of $Q$ for the pair annihilation process. The
order $\alpha$ QED corrections are then considered in Section 3, where we
report the main results of our calculations of both $vacuum$ and $thermal$
contributions. We finally discuss our results and give our conclusions in
Section 4.

\section{Pair annihilation in Born approximation}

We first consider the pair annihilation process $e^+ + e^-\rightarrow \nu +
\ov{\nu}$ at the lowest order in perturbation theory. Denoting with
$p_{1,2} \equiv (E_{1,2},{\vec{p}}_{1,2})$, the energy-momentum of ingoing
electron and positron, and with $q_{1,2} \equiv
(\omega_{1,2},{\vec{q}}_{1,2})$ the four-momentum of the outgoing neutrino
and antineutrino, respectively, the neutrino energy loss rate $Q$ is given
by\footnote{We use natural units, $\hbar = c = k =1$.}
\bea
Q &=& \frac{1}{(2 \pi)^6} \int \frac{d^3 {\vec{p}}_1}{2 E_1} \int \frac{d^3
{\vec{p}}_2}{2 E_2} \,(E_1 + E_2) \, F_-(E_1) \, F_+(E_2)  \nonumber \\
& {\times} & \left\{ \frac{1}{(2 \pi)^2}\int \frac{d^3{\vec{q}}_1}{ 2
\omega_1} \int \frac{d^3{\vec{q}}_2}{2 \omega_2} \,
\delta^4(p_1+p_2-q_1-q_2) \,     \sum_{\lambda,\beta} {|M_{e^+ e^-
\rightarrow \nu_\beta \ov{\nu}_\beta}|}^2 \right\} \, \, , \label{Q}
\eea
where $F_{\pm}(E) = \left[ \exp \left\{ \frac{E}{T} {\pm} \xi_e \right\} + 1
\right]^{-1}$ are the Fermi-Dirac distribution functions for
positrons/electrons with temperature $T$ and degeneracy parameter $\xi_e$.
Finally $\sum_{\lambda,\beta} {|M_{e^+ e^- \rightarrow \nu_\beta
\ov{\nu}_\beta}|}^2$ is the squared modulus of the reaction amplitude
summed over all particle polarizations $\lambda$ and neutrino flavours
$\beta$.

In the Born approximation, i.e. in the limit of a four fermion electroweak
interaction and no electromagnetic radiative corrections (see Fig. 1a), the
quantity in curly brackets in Eq. (\ref{Q}) takes the form
\be
\frac{2 G_F^2}{3 \pi} \left[ \left( C_V^{\prime 2} + C_A^{\prime 2} \right)
\left( m_e^2 + 2 p_1 {\cdot} p_2 \right) + 3 \left( C_V^{\prime 2} -
C_A^{\prime 2} \right) m_e^2 \right] (p_1 + p_2)^2\, \, , \label{Bexpr}
\ee
where we have defined $C_{V,A}^{\prime 2} \equiv  (1 + C_{V,A})^2 + 2
C_{V,A}^2$ with $C_V=2 \sin^2 \theta_W - 1/2$, $C_A=-1/2$.

\begin{figure}
\begin{center}
\epsfysize=10cm
\epsfxsize=14cm
\epsffile{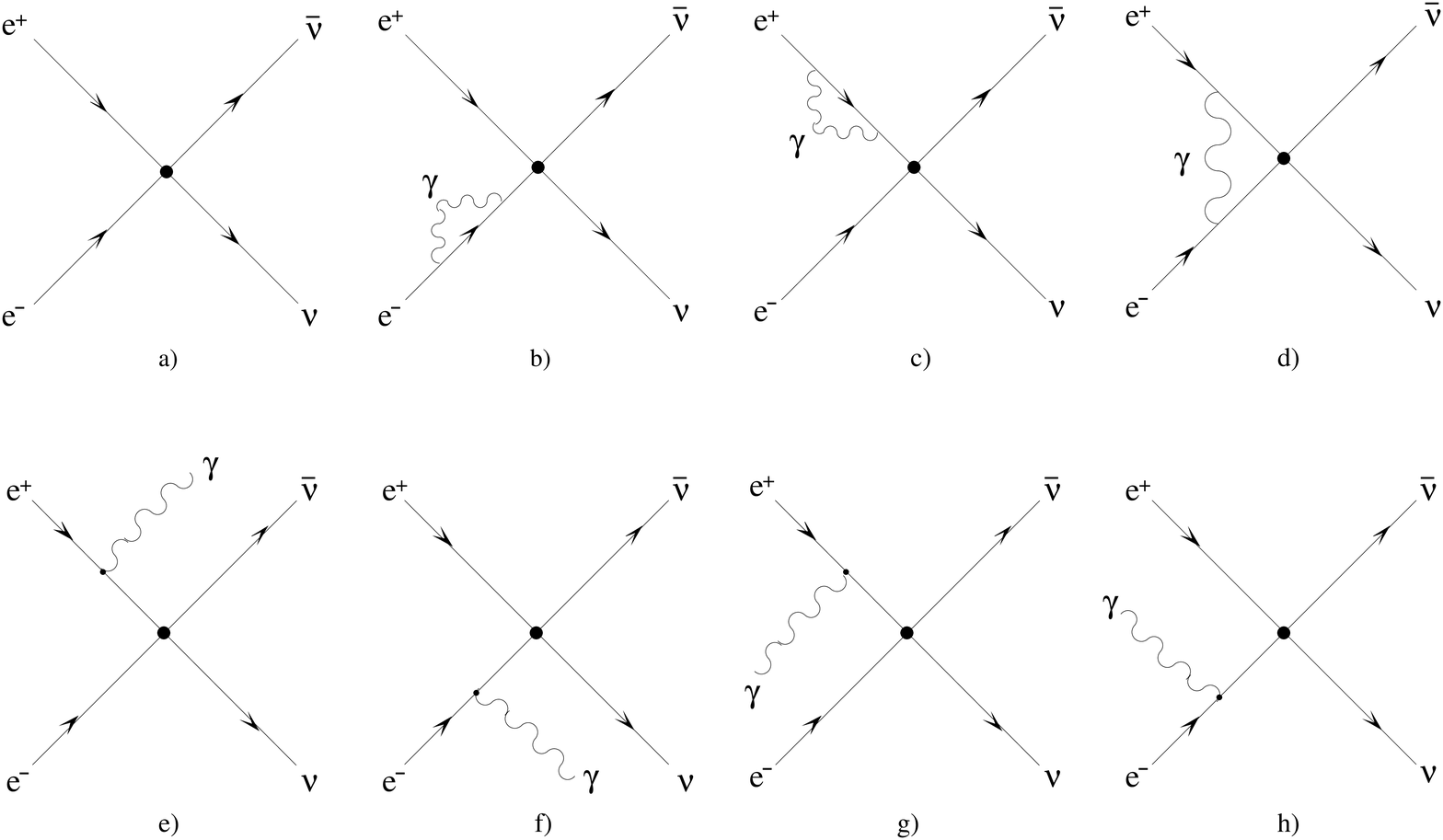}
\caption{Feynman diagrams for the pair annihilation process up to order
$\alpha G_F^2$.}
\label{pair}
\end{center}
\end{figure}

Substituting Eq. (\ref{Bexpr}) into Eq. (\ref{Q}) and performing the
angular integrations, we get the final expression for the energy loss rate
induced by pair annihilation in the Born approximation,
\bea
&Q_B & =\frac{G_F^2\, m_e^4}{18 \pi^5} \int_0^\infty \frac{|\vec{p}_1|^2 \,
d |\vec{p}_1|}{E_1} \int_0^\infty \frac{|\vec{p}_2|^2 \, d
|\vec{p}_2|}{E_2}
 \, \left(E_1+E_2\right) F_-(E_1) F_+(E_2) \nonumber \\
&{\times}&\left[ C_V^{\prime 2} \left( 7 + 9 \frac{E_1 E_2}{m_e^2}  -
\frac{E_1^2+E_2^2}{m_e^2} + 4
\frac{E_1^2 E_2^2}{m_e^4} \right) \right. \nonumber \\
 &+& \left. C_A^{\prime 2} \left( -4 -\frac{E_1^2 +E_2^2}{m_e^2} + 4
\frac{E_1^2 E_2^2}{m_e^4} \right) \right]  . \label{qpair}
\eea
Note that $Q_B$ depends on the temperature $T$ and the electron degeneracy
parameter $\xi_e$ only. It is customary to express the dependence on
$\xi_e$ (or the electron chemical potential) in terms of the matter
density, $\rho$, the electron molecular weight, $\mu_e$, and $T$ through
the condition of plasma electrical neutrality,
\begin{eqnarray}
\frac{\rho}{\mu_e} &&= \frac{1}{\pi^2 N_A} \int_{m_e}^\infty E \sqrt{E^2-m_e^2}
\, d E \, \left( \frac{1}{\exp \left\{ \frac{E}{T} - \xi_e \right\} +
1}- \frac{1}{\exp \left\{ \frac{E}{T} + \xi_e \right\} + 1} \right)~~~,
\nonumber \\
\label{rhomue}
\end{eqnarray}
$N_A$ being the Avogadro number.

\begin{figure}
\begin{center}
\epsfysize=8cm
\epsfxsize=12cm
\epsffile{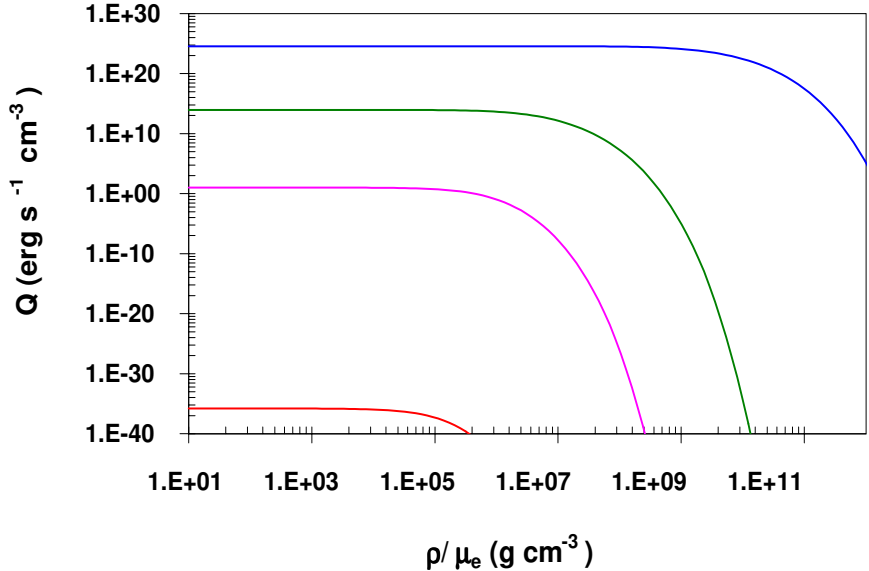}
\caption{The energy loss rate due to pair annihilation process in the Born
approximation for $T=10^8,10^{8.5},10^9,10^{10} ~^\circ K$ (from bottom to
top).}
\label{born}
\end{center}
\end{figure}

In Fig. \ref{born} we show the energy loss rate due to pair annihilation in
the Born approximation for several values of the temperature as function of
$\rho/\mu_e$.

\section{Pair annihilation: radiative corrections}

A consistent computation up to order $\alpha$ of the energy loss rate $Q$
due to pair annihilation is obtained by considering the additional
contribution of all diagrams shown in Fig.s 1b-1h. Notice that since the
energy carried by the outgoing neutrino pair is typically smaller or, at
most, of the order of 1 MeV, we can safely neglect the electroweak
radiative corrections to the four fermion effective interactions, and only
consider the set of gauge invariant purely QED contributions.

The radiative corrections in case of large values of temperature and
density must be computed taking into account the presence of the
electromagnetic plasma, namely $e^{\pm}$ and $\gamma$. As long as neutrino
mean free path is so large that they leave the star without any further
interaction, there is no relevant neutrino component in the stellar plasma.
The zero temperature radiative corrections, or {\it vacuum radiative
corrections}, have been already computed in literature. They can be
obtained for example, from the results of Ref. \cite{Passera01} by using
crossing symmetry. It is customary to classify them as follows:
\begin{itemize}
\item[-] electron mass and wavefunction renormalization (Fig.s 1b-1c);
\item[-] electromagnetic vertex correction (Fig. 1d);
\item[-] $\gamma$ emission/absorption (Fig.s 1e-1h).
\end{itemize}
The whole radiative corrections, including the finite temperature effects,
can be then obtained from the diagrams of Fig.s 1b-1h by using the real
time formalism for finite temperature quantum field theory. In this
framework the thermal propagators for electrons and photons read as follow
\bea
S_T(p) &=& \left( {\not{p}} + m \right) \Bigr\{ \frac{i}{p^2 - m_e^2 + i
\epsilon} \nonumber \\
& - & 2 \pi \, \delta \left( p^2 - m_e^2 \right) \left[ F_-(|p_0|)
\Theta (p_0) + F_+(|p_0|) \Theta (-p_0) \right] \Bigr\}~~~,  \label{ProE}
\\
D_T^{\alpha \beta}(k) & = & - \left[ \frac{i}{k^2 + i \epsilon} + 2 \pi \,
\delta \left( k^2 \right) B(|k_0|) \right] g^{\alpha \beta} ~~~, \label{PF}
\eea
where $\Theta (x)$ is the step function and $B(x)$ is the Bose distribution
function. The first terms in Eq. (\ref{ProE}) and (\ref{PF}) are the usual
$T=0$ Feynman propagators, while those depending on the temperature (and
density) through the distribution functions describe the interactions with
real particles of the thermal bath.

The calculation for the order $\alpha G_F^2$ corrections to the Born
approximation for the energy loss rate proceeds in close analogy with what
has been described in Ref.s \cite{emmp98,emmp99} for a very similar
situation. In particular, the additional contributions to $Q$ due to mass,
wave function and vertex corrections come from the interference between
diagrams of Fig.s 1b, 1c and 1d with the Born amplitude of Fig. 1a. In
addition, the energy loss rates due to $\gamma$ emission (absorption)
$Q_{e(a)}$, are given by the sum of squared amplitudes of the processes of
Fig.s 1e, 1f (1g, 1h),
\bea
Q_e &=& \frac{1}{(2\pi)^9} \int \frac{d^3{\vec{p}}_1}{ 2E_1} \,
\int\frac{d^3 \vec{p}_2}{2E_2} \, \int\frac{d^3 \vec{k}}{2 \omega} \,
(E_1+E_2-\omega) F_-(E_1) \, F_+(E_2) \, (1+B(\omega)) \nonumber \\
& {\times} & \left\{ \frac{1}{(2 \pi)^2}\int \frac{d^3{\vec{q}}_1}{ 2
\omega_1}    \int \frac{d^3{\vec{q}}_2}{2 \omega_2} \,
\delta^4(p_1+p_2-q_1-q_2-k) \,     \sum_{\lambda,\beta} |M_{e^+ e^-
\rightarrow \nu_\beta \ov{\nu}_\beta \gamma}|^2 \right\}\, \, ,\nonumber \\
\label{PE} \\ Q_a &=& \frac{1}{(2\pi)^9} \int \frac{d^3 \vec{p}_1}{ 2E_1}
\,
\int\frac{d^3 \vec{p}_2}{2E_2} \, \int\frac{d^3 \vec{k}}{2 \omega} \,
(E_1+E_2+\omega) F_-(E_1) \, F_+(E_2) \, B(\omega) \nonumber \\
& {\times} & \left\{ \frac{1}{(2 \pi)^2}\int \frac{d^3{\vec{q}}_1}{ 2
\omega_1}    \int \frac{d^3{\vec{q}}_2}{2 \omega_2} \,
\delta^4(p_1+p_2-q_1-q_2+k) \,     \sum_{\lambda,\beta} |M_{e^+ e^- \gamma
\rightarrow \nu_\beta \ov{\nu}_\beta}|^2 \right\}~~~. \nonumber \\
\label{PA}
\eea
For the seek of brevity we do not report here the details of our
calculation leaving this lengthy description to a forthcoming paper
\cite{emmpp}. We only briefly outline the method adopted to obtain the
final corrections $\Delta Q$. Each of the several contribution to $|M_{e^+
e^- \rightarrow \nu_\beta \ov{\nu}_\beta}|^2$, and the two squared
amplitude $|M_{e^+ e^- \rightarrow \nu_\beta \ov{\nu}_\beta \gamma}|^2$ and
$|M_{e^+ e^- \gamma \rightarrow \nu_\beta \ov{\nu}_\beta}|^2$ have been
evaluated analytically. The results should be then integrated over the
relevant phase space. Some of these integrations can be again done
analytically, while we have used a Montecarlo technique to deal with the
(usually three of five-dimensional) remaining integrations, requiring an
accuracy better than $1 \%$ on $\Delta Q$ \footnote{The accuracy is
actually depending on the values of temperature and density. The estimate
of $1 \%$ for the theoretical error is rather conservative, and represents
the larger result in the whole interval where pair annihilation
significantly contribute to the cooling rate $Q$, $10^{9}\,^\circ K \leq T
\leq 10^{10} \,^\circ K$ . For smaller values of $T$, since the pair
annihilation process rate is very small, the accuracy gets worst, and in
our calculations can reach $5 \%$.}. The main difficulty of this procedure
is that each contribution coming from the interference of diagrams 1b)-1d),
as well as the squared amplitudes 1e)-1h) is plagued by infrared
divergencies, while the overall result is of course divergence free. To
regularize these divergencies we have explicitly subtracted all divergent
terms expanding the squared amplitudes in a Laurent series around the pole
singularities. This method follows quite closely what has been already used
in \cite{emmp98,emmp99,camb}.

\begin{figure}
\begin{center}
\epsfysize=8cm
\epsfxsize=12cm
\epsffile{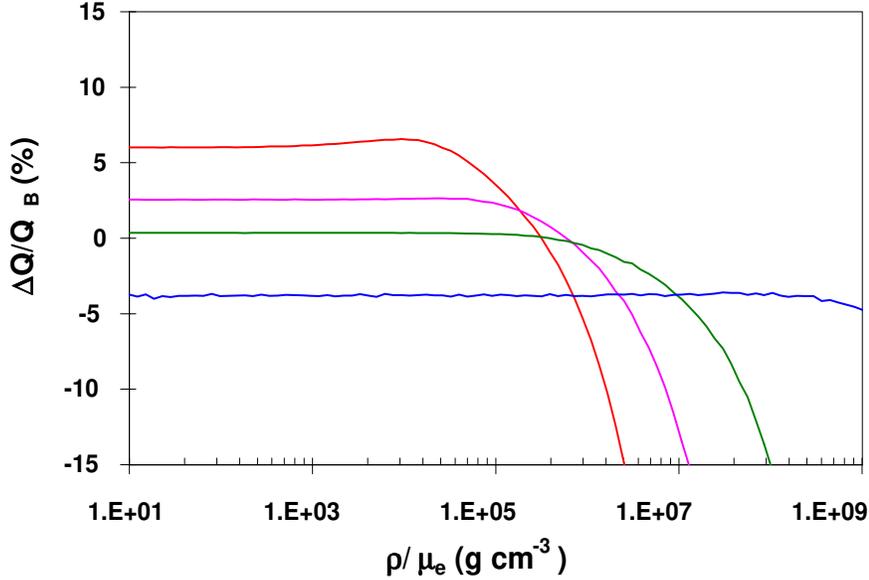}
\caption{The total radiative corrections normalized to the Born
approximation result for the pair annihilation process for
$T=10^8,10^{8.5},10^9,10^{10} ~^\circ K$ (from top to bottom).}
\label{paircorr}
\end{center}
\end{figure}

In Fig. \ref{paircorr} we show the ratio between the sum of all radiative
corrections and $Q_B$, $\Delta Q/Q_B$, for some values of $T$ as a function
of the density. For a fixed value of temperature, the ratio $\Delta Q/Q_B$,
shows a plateau, for all densities for which the electron-positron chemical
potential is smaller than the temperature, so that the mean energy is
essentially given by $T$. The plateau end point, as expected, shifts
towards larger values of $\rho_{\mu_e}$ with increasing temperature. For
even larger densities the mean energy of the $e^+-e^-$ pair is dominated by
their Fermi energy $\epsilon_F$, and the ratio $\Delta Q/Q_B$ start varying
logarithmically as a function of $\epsilon_F/m_e$. Similarly, at fixed
density and increasing temperature, where the effect of degeneracy can be
neglected, the ratio $\Delta Q/Q_B$ shows a similar logarithmic behaviour
as function of $T/m_e$\footnote{See, for example, the typical behaviour of
QED radiative corrections for a similar process in \cite{Passera01}}. Both
these behaviours suggest that in these extreme regimes higher order QED
contributions should be included. However, the trends exhibited in Fig.
\ref{paircorr} cannot be straightforwardly extrapolated where $T$ or
$\epsilon_F $ start to be closer to the $W$ and $Z$ masses, since the
effects of electroweak corrections should be taken into account as well.
Furthermore it is also worth noticing that for temperatures larger than
$10^{10} \, ^\circ K$, and/or densities larger than, say $10^{12}$ g
cm$^{-3}$, the neutrino mean free path becomes of the order or smaller than
the dimension of the star core, so that neutrinos start interacting with
the medium. This means that to evaluate the cooling rate via neutrino
emission it is necessary to consider their Pauli blocking factors, as well
as the (negative) contribution due to inverse processes.

It is actually interesting to see how the increasing importance of
radiative corrections with temperature and density in the range of interest
emerges from the study of other parameters which characterise the
properties of the electromagnetic plasma, like thermal effective masses of
particles. In general, thermal masses would show an explicit dependence on
the spatial momentum $\vec{p}$. Nevertheless a nice way to see how their
dispersion relations are influenced by interactions is to consider their
average over the equilibrium distribution. In particular, for the electron
effective mass we have
\be
<m_e^R> = \frac{1}{\pi^2 \, n_-}\int_0^\infty |\vec{p}|^2 \, d {|\vec{p}|}\,
\, m_e^R(|\vec{p}|) \, \, F_-(E) ~~~, \label{me}
\ee
where $n_-$ is the electron number density and $m_e^R(|\vec{p}|)$ is the
momentum-dependent renormalized mass in the electromagnetic plasma with
temperature $T$ and density $\rho/\mu_e$,
\bea
 \!\!\!\!\!\!\!\!\!\!\!\!m_e^R(|\vec{p}|) = m_e + \frac{\alpha \pi}{3}
\frac{T^2}{m_e} +  \frac{\alpha}{\pi m_e} \int_0^\infty \frac{|\vec{k}|^2
\, d|\vec{k}|}{\omega} \left( F_-(\omega) + F_+(\omega) \right) -
\frac{\alpha m_e}{2 \pi |\vec{p}|} \int_0^\infty \frac{|\vec{k}| \,
d|\vec{k}|}{\omega} \nonumber \\
{\times} \left( F_-(\omega) \log \left(\frac{\omega E - m_e^2 + |\vec{p}|
|\vec{k}|}{\omega E - m_e^2 - |\vec{p}| |\vec{k}|} \right) - F_+(\omega)
\log \left(\frac{\omega E + m_e^2 + |\vec{p}| |\vec{k}|}{\omega E + m_e^2 -
|\vec{p}| |\vec{k}|}\right) \right), \,\,\,\,\,\,\, \nonumber\\ \label{meme}
\eea
where $\omega=\sqrt{|\vec{k}|^2+m_e^2}$. An analogous result for positron
can be obtained from (\ref{meme}) with the substitutions $F_{\pm}(\omega)
\rightarrow F_\mp(\omega)$.

\begin{figure}
\begin{center}
\epsfysize=8.5cm
\epsfxsize=8.5cm
\epsffile{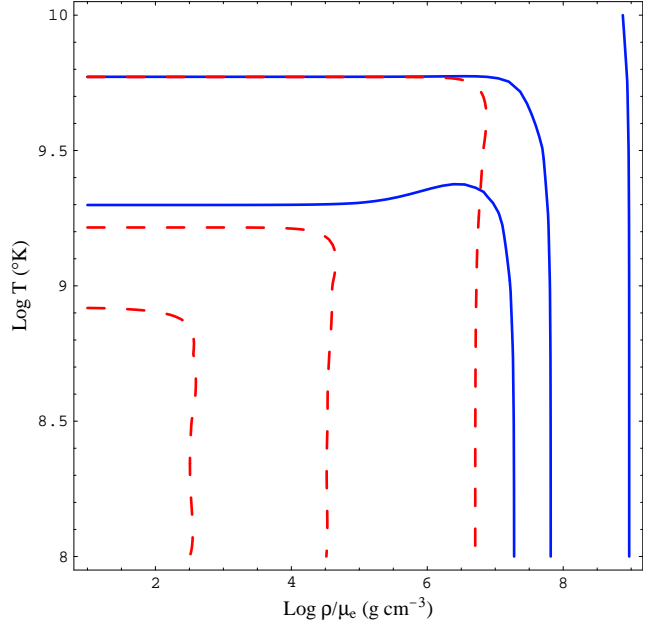}
\caption{Contour plots for the photon (dashed lines) and the electron
(solid lines) thermal mass correction, normalized to $m_e$. The electron
effective mass has been averaged over the thermal equilibrium distribution.
From left to right the curves refer to the values $10^{-3}$, $10^{-2}$, and
$10^{-1}$.}
\label{mteg}
\end{center}
\end{figure}

For the photon thermal mass the averaging is unnecessary, since it does not
depend on the spatial momentum
\be
m_\gamma^2 = \frac{4 \alpha}{\pi} \int_0^\infty \frac{|\vec{k}|^2 \,
d|\vec{k}|}{\omega} \left( F_-(\omega) + F_+(\omega) \right)\,\,\, .
\label{mp}
\ee
In Fig. \ref{mteg} we have reported the contour plots of
$(<m_e^R>-\,\,m_e)/m_e$ and $m_\gamma/m_e$ in the temperature-density
plane. Curves from left to right refer to the values $10^{-3}$, $10^{-2}$
and $10^{-1}$, respectively. From this plot it is clear that both electron
and photon effective masses grows logarithmically with the particle mean
energy, and are again of the order of few percent in the relevant range of
temperature and density.

\section{Results and conclusions}

In this letter we have reported on the calculation of the energy loss rate
of pair annihilation process i) up to order $\alpha G_F^2$. The Born value
for $Q$ has been corrected including both $vacuum$ and $thermal$ radiative
corrections, the latter being computed in the real time formalism. One of
our main results is presented in Fig. \ref{paircorr}, where we plot these
radiative corrections with respect to the Born approximation estimate, as
functions of the plasma density for some values of the temperature. The
corrections come out to be of the order of few percent and negative for
high temperatures, implying that for these temperatures the energy loss is
sensibly decreased. At fixed temperature, $\Delta Q / Q_B$ goes to a
constant value for low density. This can be easily understood, since in
this limit the plasma is weakly degenerate, and therefore the energy loss
rate depends on temperature only. At large densities the ratio $\Delta
Q/Q_B$ decreases and reach larger negative values. However for such high
densities the pair annihilation rates are exceedingly small and thus this
process gives only a marginal contribution to the star cooling.

\begin{figure}
\begin{center}
\begin{tabular}{ll}
\epsfxsize=7cm
\epsfysize=10cm
\epsffile{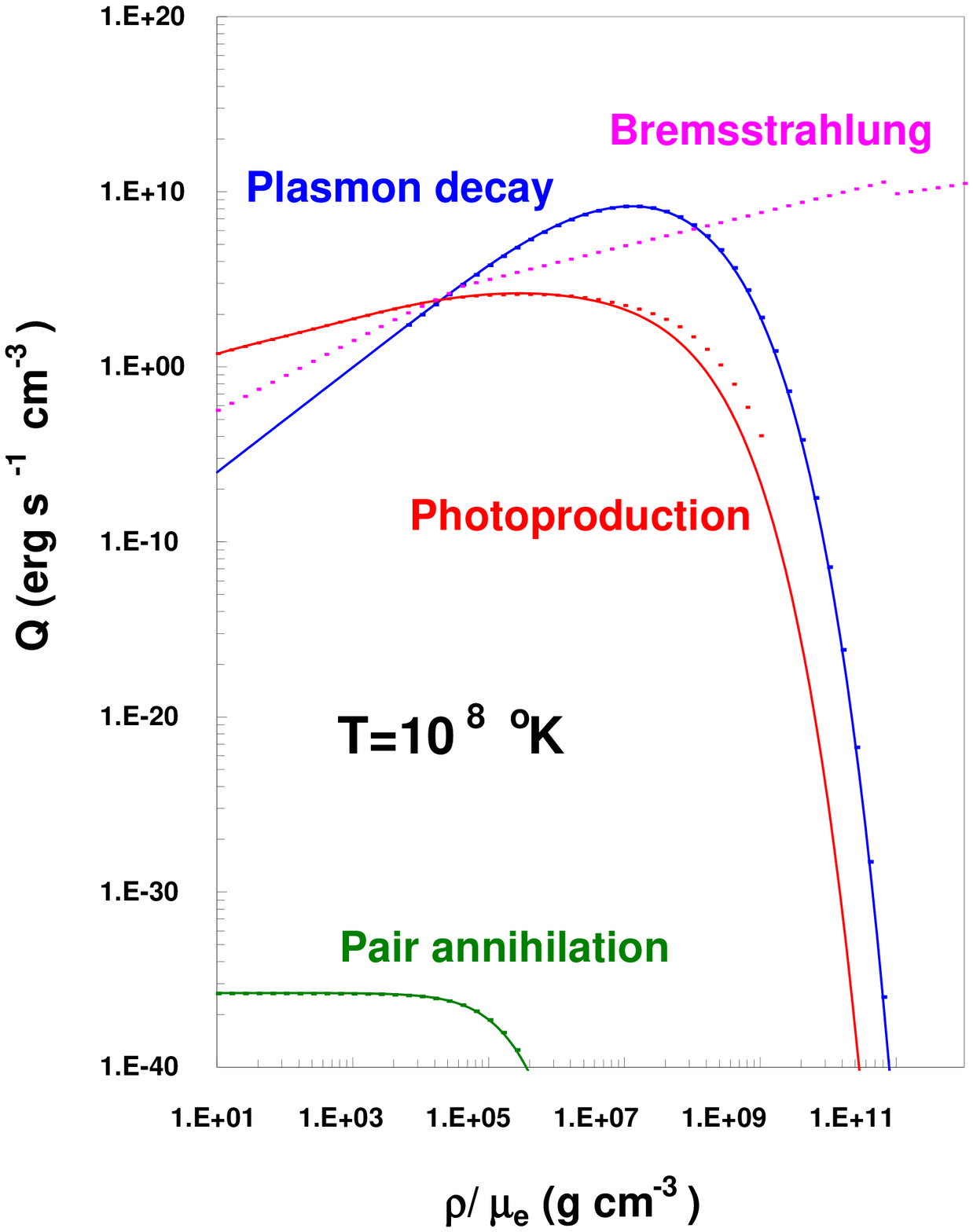}
&
\epsfxsize=7cm
\epsfysize=10cm
\epsffile{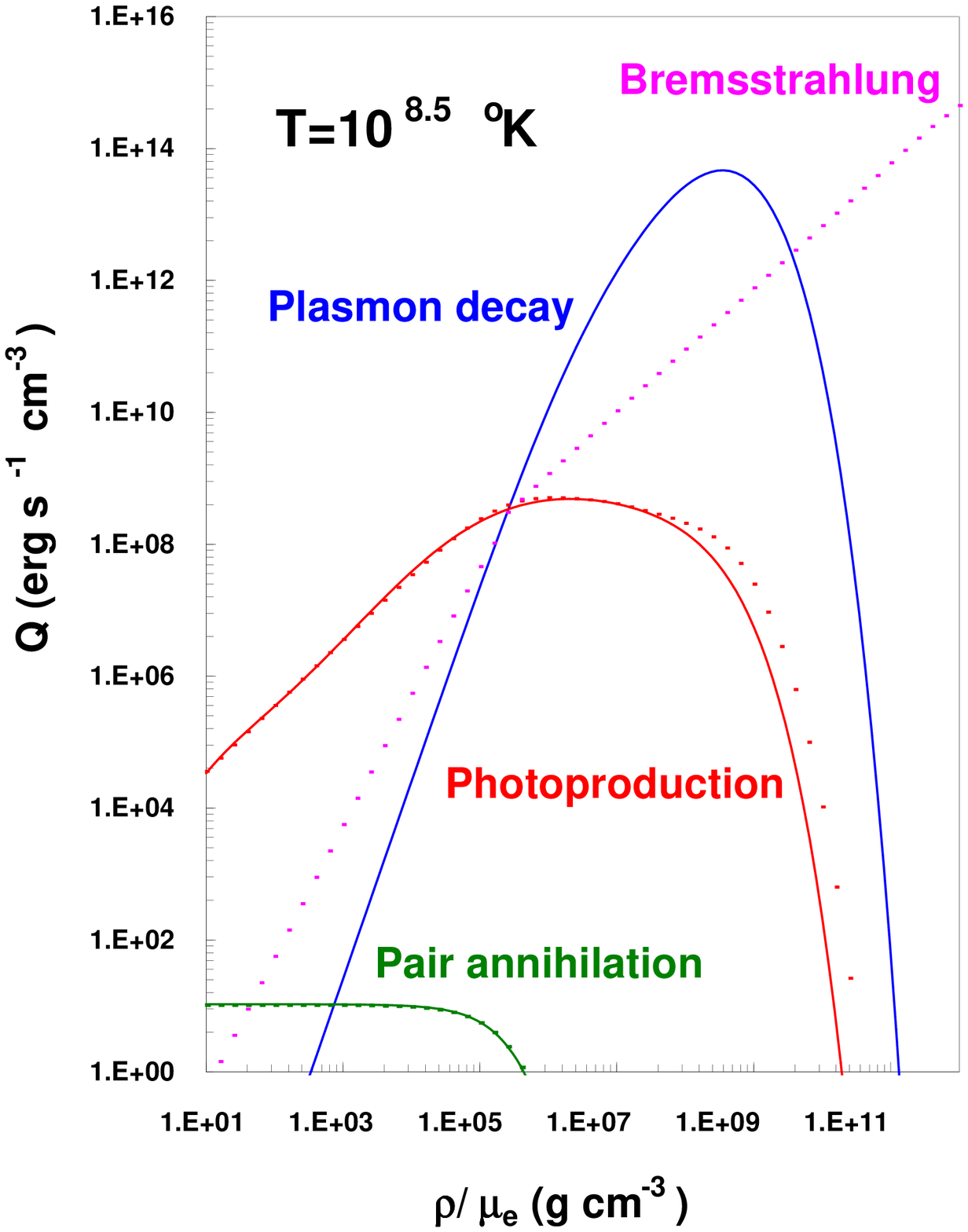}
\\
\epsfxsize=7cm
\epsfysize=10cm
\epsffile{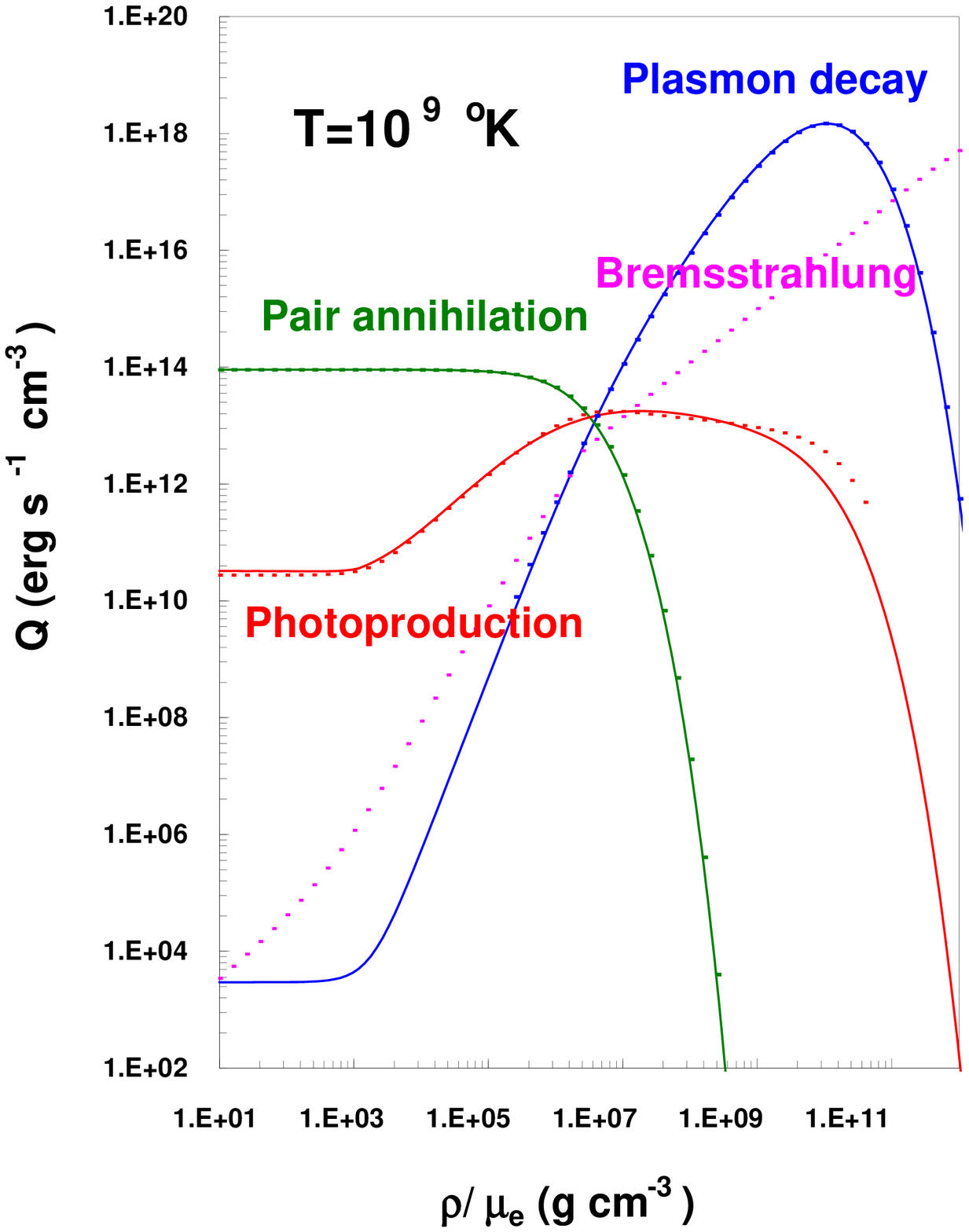}
&
\epsfxsize=7cm
\epsfysize=10cm
\epsffile{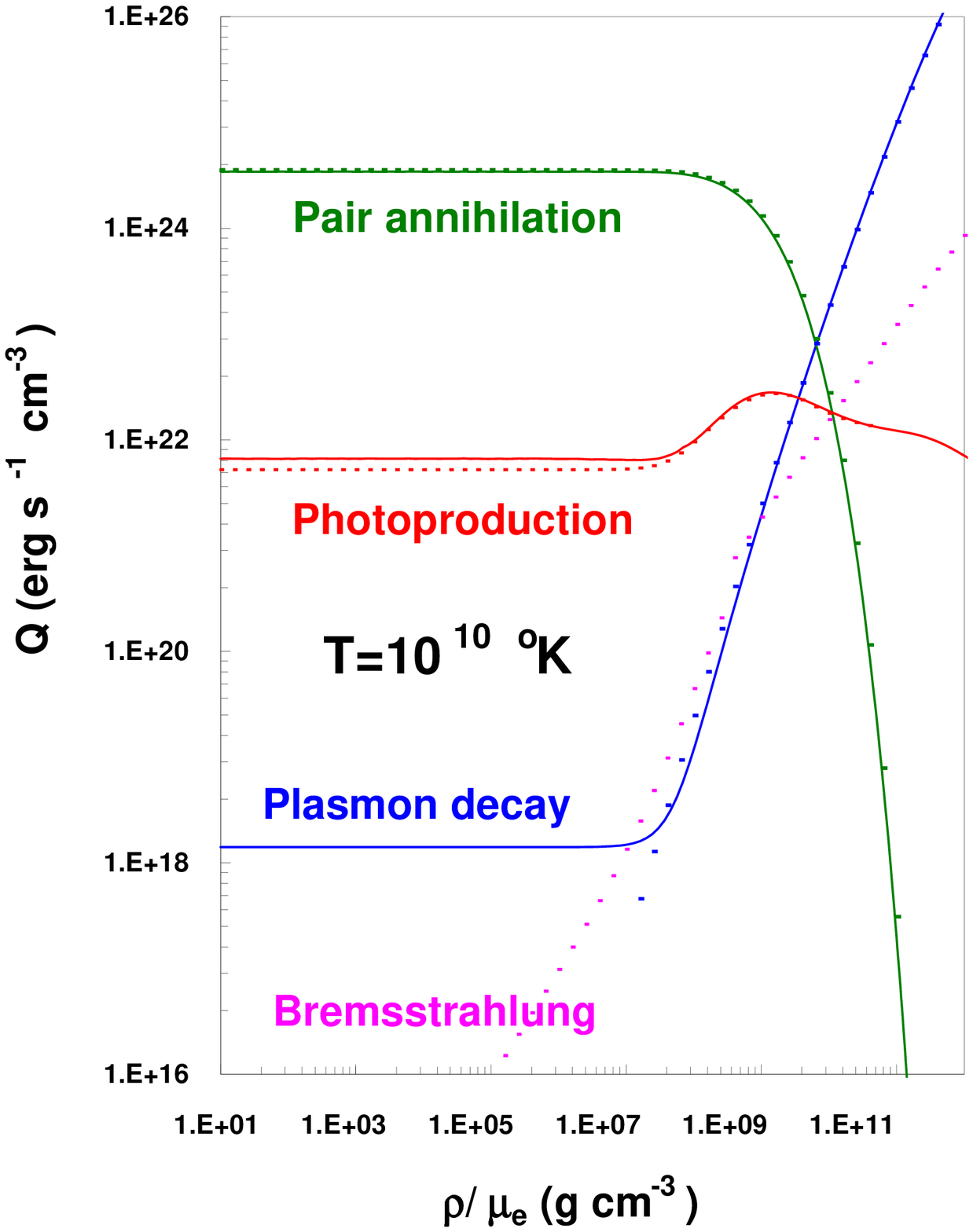}
\end{tabular}
\end{center}
\caption{The energy loss rate versus $\rho/\mu_e$ due to pair annihilation
(including radiative corrections), photoproduction and plasmon decay (solid
lines) for several temperatures. The dotted lines refer to the analogous
results of Ref. \cite{Itoh96}, which also compute the rate for
bremsstrahlung on nuclei. The effect of $\Delta Q$ to pair annihilation can
be appreciated in this logarithmic scale only for the largest temperature
$10^{10} \,^\circ K$.}
\label{qcomp}
\end{figure}

In order to perform a comprehensive study of order $\alpha G_F^2$
contributions to the energy loss rate, we have also recalculated the
contribution from $\nu$-photo\-production and plasmon decay processes.
While the detailed analytical calculations will be discussed elsewhere
\cite{emmpp}, we here give our numerical results in Fig. \ref{qcomp}, where
we show the several contributions to $Q$. When possible, we also show for
comparison the results of Ref. \cite{Itoh96}. In particular our results for
these processes agree quite nicely, at least in the whole region where they
significantly contribute to the total energy loss rate. For the
bremsstrahlung on nuclei we have used the analytic fitting formula of Ref.
\cite{Itoh96}.

Fig. \ref{qcomp} shows that, as well known, for large temperatures and not
too high densities, pair annihilation dominates over the other two
processes, while for low densities $\nu$-photoproduction dominates over
plasmon decay. On the other hand, for large densities the most relevant
process is plasmon decay, whose rate however, along with those of all other
processes, rapidly falls down for extremely high densities. This is a
genuine plasma effect. Consider, for example, the behaviour of
$\nu$-photoproduction energy loss. As already noted in \cite{Beaudet67b},
the decrease for very large densities is achieved only if one consistently
takes into account the increasingly large photon thermal mass. In fact with
a massless photon the $\nu$-photoproduction curves in Fig. \ref{qcomp}
would rather reach a constant value. The main effect of $m_\gamma^2$ is a
lowering of the values of the Bose distribution function for photons, i.e.
a smaller number of thermal photons. This reduces the energy loss rate
induced by $\nu$-photoproduction.

\begin{figure}
\begin{center}
\epsfysize=8.5cm
\epsfxsize=8.5cm
\epsffile{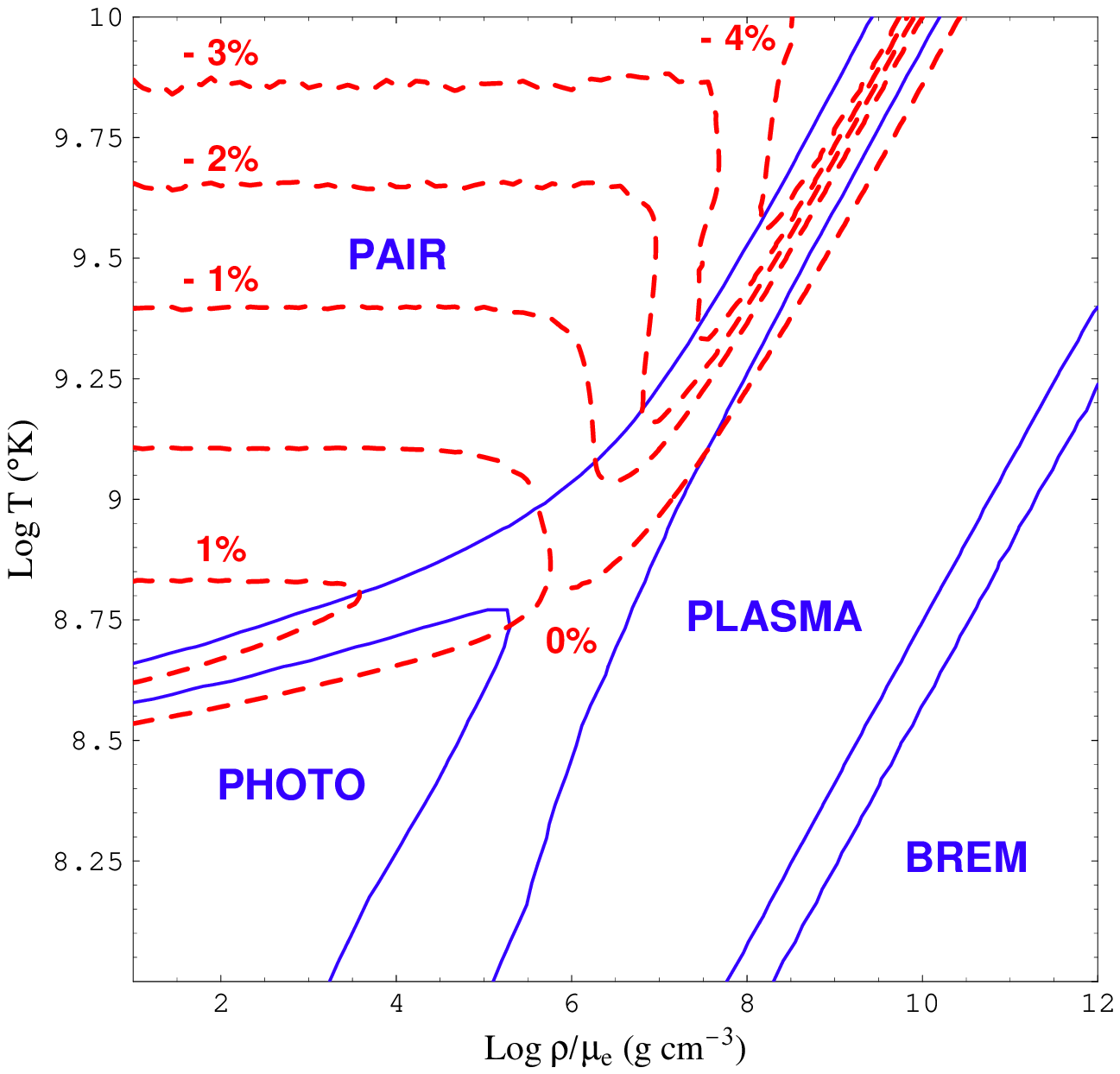}
\caption{The regions in the $T-\rho/\mu_e$ plane where each of the
processes i)-iv) contribute for more than $90 \%$ to the total energy loss
rate. We also show the contours for the relative correction $\Delta
Q/Q^0_{Tot}$ (see text) for the values $1\%,0\%,-1\%,-2\%,-3\%,-4 \%$.}
\label{qcontour}
\end{center}
\end{figure}

In Fig. \ref{qcontour} we show the regions in the temperature-density plane
where a given process contributes to the total energy loss rate (including
radiative corrections to pair annihilation) for more than 90\%. We also
summarize there our results on the radiative corrections to pair
annihilation processes, by plotting the contours corresponding to $\Delta
Q/Q_{Tot}^0=1\%,0\%,-1\%,-2\%,-3\%,-4 \%$, where $Q_{Tot}^0$ is the total
emission rate with pair annihilation calculated in Born approximation.
These contours lie almost entirely in the region where the pair
annihilation process gives the main contribution to $Q$. This result may
affect the late stages of evolution of very massive stars by changing their
configuration at the onset of Supernova explosion. This issue is presently
under study.

\section{Acknowledgements}
We warmly thank M. Passera for a clarifying correspondence and suggestions.
We are also pleased to thank A. Chieffi, G.Imbriani, M. Limongi, L.
Piersanti and O. Straniero for useful discussions.

\end{document}